# Securing Web Services Using XML Signature and XML Encryption

**RA. K. Saravanaguru[1], George Abraham[2], Krishnakumar Venkatasubramanian[3], Kiransinh Borasia[4]**

School of Computer Science and Engineering,
VIT University, Vellore, India

[1]sarophd@gmail.com
[2]george.abraham2011@vit.ac.in
[3]venkatasubramanian2011@vit.ac.in
[4]borasiakiransinh.ranjitsinh2011@vit.ac.in

**Abstract:** This paper is aimed to evaluate the importance of XML Signature and XML Encryption for WS-Security. In today's e-business scenario, organizations are investing a huge amount of their resources in Web Services. Web Service Transactions are done mainly through plain-text XML formats like SOAP and WSDL, hence hacking them is not a tedious task. XML Signature and XML Encryption ensure security to XML documents as well as retain the structure of the documents, thereby making it easy to implement them. These two methods are evaluated on the parameters of authentication, authorization, integration, confidentiality and non-repudiation.

## 1. Introduction

The concept of Web Services started in the late 1990's and from then on has become the backbone of the IT industry. Currently all business transactions rely on Web Services to achieve their desired goals. With the portability and customizability of Extensible Markup Language (XML), it has become the universal language for all these Web Services (XML and Web Service – Unleashed; XML for Dummies). XML Web Services are a successful paradigm for many complex web-based applications (Sun and Li, 2005). Web Services use XML as an interface or medium to define business functions and communicate with each other.

Due to its wide usage, web service security is of growing demand day by day. For a successful business environment, it is a necessity that all the applications and communications must be secure and reliable. In the present world, Internet is the widely used medium for all business transactions and therefore, security is the main concern. In this paper, we would be first giving an overview of Web Service Security followed by a description on the WS-Security Architecture. Then we talk about XML Signature, its types and syntax followed by an overview on XML Encryption and its syntax. Next we have a detailed description of the existing work that has been done in the field of web service security using XML Signature and XML Encryption followed by a comparative summarization of the various facilities offered by both.

## 2. Web Service Security

Web Services, like common web applications, relies on the same HTTP transport protocol and the basic web architecture. Hence it is susceptible to similar threats and vulnerabilities.

Web Service Security (WS-Security) is a flexible and feature-rich extension to SOAP to apply security to web services. It is a member of the WS-* family of web service specifications and was published by OASIS (Web Services Security – Wikipedia).

Some of the basic concepts that Web Services Security are based upon are (Bertino, Carminati and Ferrari, 2001; Han, Park and Lim, 2011; Nordbotten, 2009; Singhal, Winograd and Scarfone, 2007; Web Services Security – Wikipedia):

1. Identification and Authentication: Verifying the identity of the user, process or device to allow access to a resource or information system
2. Authorization: The permission to use a resource
3. Integrity: The property that the data has not been modified in any unauthorized manner while in storage, processing or transit

4 Non-repudiation: Non-denial by either sender or receiver of having sent or received the information, respectively
5 Confidentiality: Preserving authorized restriction and information access
6 Privacy: Restricting access to subscriber or relying party information in accordance with Federal Law and organizational policy

*2.1 WS-Security Architecture*

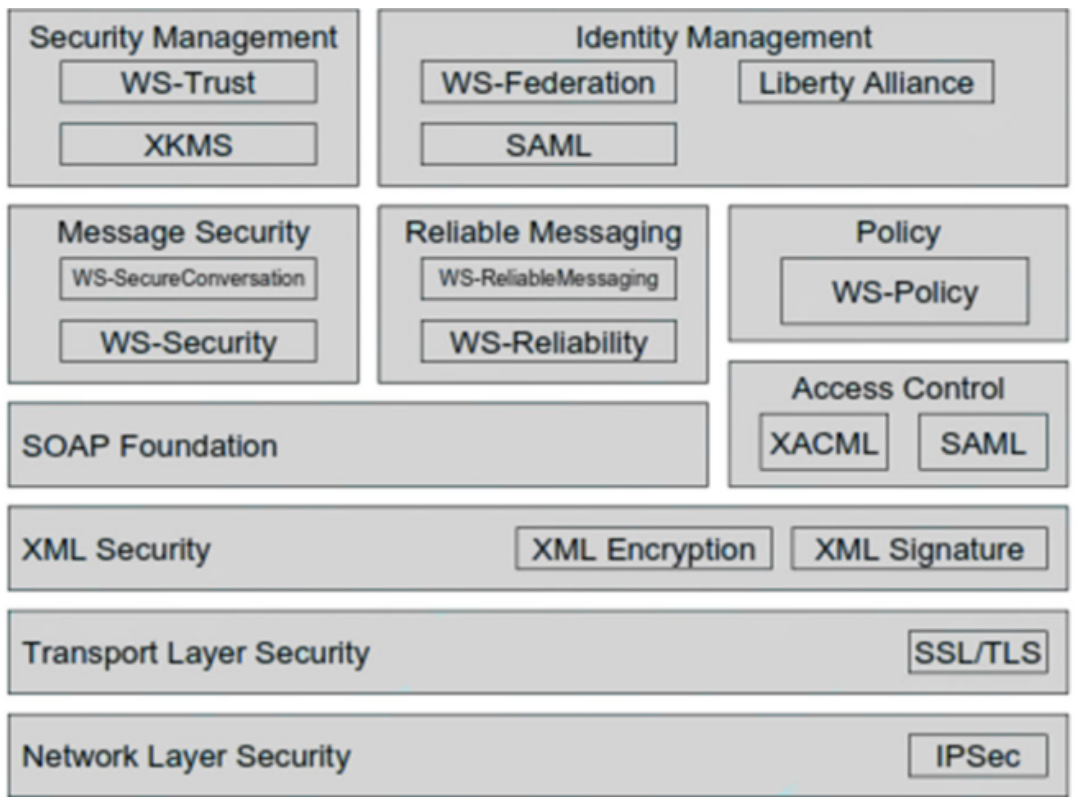

Fig. 1 WS-Security Architecture
(Singhal, Winograd and Scarfone, 2007)

The open community that created Web Services developed a number of security standards for Web Services. The above reference model maps these standards to the various layers of the standard Web Service (Singhal, Winograd and Scarfone, 2007):

1 WS-Trust: Describes a framework for trust models that enables Web Services to operate securely
2 WS-Policy: Describes the capabilities and constraints of the security policies on intermediaries and endpoints
3 WS-Privacy: Describes a model for how Web Services and requesters state privacy preferences and organizational privacy practice statements
4 WS-Security: Describes how to attach signatures and encryption headers to SOAP messages
5 WS-Federation: Describe how to manage and broker the trust relationships in a heterogeneous federated environment including support for federated identities
6 WS-SecureConversation (Nordbotten, 2009): Describe how to manage and authenticate message exchanges between parties including security context exchange and establishing and deriving session keys

There are two standards for XML Security – XML Signature and XML Encryption. For our discussion, we will be concentrating mainly on XML Signature for XML Security, explaining the basic concept and its various implementation techniques and finally summarizing with a comparison with XML Encryption.

## 3. XML Signature

Digital Signatures have become an important aspect of electronic security because they can be used to ensure the integrity, authenticity and non-repudiation of data (Sun and Li, 2005). XML Signature, is a W3C recommendation, released on 12th February, 2002 (XML Digital Signature – www.w3.org), in which the digital signatures are optimized for XML documents, for ensuring integrity of XML data during signing and/or verification process (Knap and Mlynkova, 2009).

The practical benefits of XML Signature include Partial Signature, which allows signing of specific tags contained in XML data, and Multiple Signature, which allows signing multiple tags in XML document. The use of XML Signature can solve security problems, including falsification, spoofing, and repudiation.

The XML Signature supports any type of digital signature encryption using all possible standard encryption algorithms. The XML Signature does not represent the primary data, but the encrypted data in the document. MD5, SHA-1 and RSA are some of the algorithms that are used to calculate the hash values of the data. The signature process is carried out on these hash values. The hash value is called as the "fingerprint" of the primary data and any small change in the primary data will lead to a large change in the hash value due to "avalanche effect" (Yue-sheng, Meng-tao and Yong, 2010). After the hash value is signed, it is guaranteed that the integrity of the original document cannot be changed (Yue-sheng, Meng-tao and Yong, 2010).

*3.1 Types of XML Signature*

There are three basic types of XML Signatures (Bertino, Carminati and Ferrari, 2001; Han, Park and Lim, 2011; Nordbotten, 2009; Sun and Li, 2005): Enveloped Signature, Enveloping Signature and Detached Signature.

In Enveloped XML Signature, as shown in figure 2, the XML Signature is included in the document itself and is the child element of the object being signed. The data being signed envelopes the <signature> and </signature> tags. In Enveloping XML Signature, as shown in figure 3, the document is included in the XML Signature as the child element of the XML Signature. The data being signed is enclosed in the <signature> and </signature> tags. In Detached XML Signature, as shown in figure 4, the XML Signature is a separate document (mainly non-XML) from the signed XML document. The location of the signed document is given as a reference in the XML Signature.

```
<document>
     …… Data
   <signature>
     …… Contains
     reference to   the
     Data being signed
   </signature>
</document>
```

Fig. 2 Enveloped XML Signature

```
<signature>
     …… Contains
     reference to the
     Data being signed
   <document>
     …… Data
   </document>
</signature>
```

Fig. 3 Enveloping XML Signature

```
<signature>
     …… Contains
     reference to the
     signed data
</signature>
```

Fig. 4 Detached XML Signature

*3.2 XML Signature Structure*

XML digital signatures are represented by the Signature element which has the following structure, where "?" denotes zero or one occurrence; "+" denotes one or more occurrences; and "*" denotes zero or more occurrences. (XML Digital Signature – Recommendation – www.w3.org; Yue-sheng, Bao-jian and Wu, 2009)

The Signature element contains three main information (Bertino, Carminati and Ferrari, 2001; Han, Park and Lim, 2011; Knap and Mlynkova, 2009; Nordbootten, 2009; XML Signature – Wikipedia; Yue-sheng, Bao-jian and Wu, 2009):

1 SignedInfo
This contains information about the signed collection of XML fragments. This contains or references the signed data and specifies what algorithms are used. CanonicalizationMethod element specifies an algorithm that is used for normalization of the SignedInfo element before it is actually signed by an algorithm mentioned in the element SignatureMethod. Multiple Reference elements can be present. Each Reference element describes one hashed XML fragment where – Transforms element contains an arbitrary number of elements Transform which depicts a set of transforms that must be performed on the XML fragment, the method for hashing the XML fragment is specified in the DigestMethod element and the hashed value of the transformed elements is stored in the element DigestValue.

2 SignatureValue
After signing the XML document with the algorithm mentioned in the CanonicalizationMethod element, signature is generated with the parameters specified in the SignatureMethod element and the result is stored in the SignatureValue element.

3 KeyInfo
This is an optional element that is used by the signer to provide the recipients with the appropriate key that validates the signature, usually in the form of one or more X.509 digital certificates.
The Object element is an optional field that contains the data that is signed, as it is an enveloping signature type.

```
<Signature ID?>
   <SignedInfo>
      <CanonicalizationMethod/>
      <SignatureMethod/>
      (<Reference URI? >
         (<Transforms>)?
         <DigestMethod>
         <DigestValue>
      </Reference>)+
   </SignedInfo>
   <SignatureValue>
   (<KeyInfo>)?
   (<Object ID?>)*
</Signature>
```

Fig. 5 XML Signature Structure
(Bertino, Carminati and Ferrari, 2001; XML Signature – Wikipedia; Yue-sheng, Bao-jian and Wu, 2009)

## 4. Brief Overview of XML Encryption

Encryption is the method of conversion of the sensitive document into a form that is not understandable to unauthorized users. Authorized users have to decrypt the ciphered text in order to understand the content. It is a very old technique to achieve data security. There are various standard encryption algorithms available, like symmetric key encryption and asymmetric (public) key encryption, and these can also be applied to the normal XML documents. This process is called as XML Encryption. XML Encryption is a W3C Recommendation released on 10th December, 2002.

*4.1 XML Encryption Syntax*

XML Encryption is represented by the EncryptedData (XML Encryption – Wikipedia; XML Encryption – www.w3.org) element which has the following structure, where "?" denotes zero or one occurrence; "+" denotes one or more occurrences; "*" denotes zero or more occurrences; and the empty element tag means the element must be empty (XML Encryption – www.w3.org).

XML Encryption and XML Signature, both use KeyInfo element which provides information to a recipient about what keying material to use in validating a signature or decrypting encrypted data. The CipherData element contains the ciphered value and also reference to the data that is to be ciphered.

```
<EncryptedData Id? Type? MimeType? Encoding?>
  <EncryptionMethod/>?
  <ds:KeyInfo>
    <EncryptedKey>?
    <AgreementMethod>?
    <ds:KeyName>?
    <ds:RetrievalMethod>?
    <ds:*>?
  </ds:KeyInfo>?
  <CipherData>
    <CipherValue>?
    <CipherReference URI?>?
  </CipherData>
  <EncryptionProperties>?
</EncryptedData>
```

Fig. 6 XML Encryption (XML Encryption - Wikipedia)

## 5. Existing Work

In the area of Web Service Security, XML Signature was the first W3C Recommendation released, which aided the use of digital signature for signing XML documents so as to attain the qualities of integrity, authentication, authorization and non-repudiation. Later on, another W3C Recommendation was released called XML Encryption, which aided the encryption of XML data with various algorithms, so as to attain the same benefits. Both XML Signature and XML Encryption support partial, multiple as well as complex signature and/or encryption.

Lautenbach (2004), in his paper, starts by talking about the need of XML Security Standards. He says that "At first glance, signing or encrypting a XML document appears to be no different to signing or encrypting any other document." Using normal encryption technology or signing methods on XML documents led to problems such as:

1. Inability of normal signatures to handle changes caused by parsing and re-serialization
2. Inability of normal signatures to handle changes in the character sets
3. Inability to easily represent the value of a signature or the output of an encrypted XML document in an XML format

He talks about the method of implementing both XML Signature and XML Encryption for XML security, their syntaxes and the various processing rules. Apart from that for XML Signature, he talks about XML Canonicalization (representing XML in the form of a canonical byte form) and the various transformations and with respect to XML Encryption, he discusses the granularization (describing what will be encrypted). He also talks about some implementation issues of XML Signatures complex XPath transforms being too CPU intensive and canonicalization of large XML document being slow. He also gives a brief idea of some related standards like XPath Filter 2 Transform, Exclusive XML Canonicalization, XML Key Management Specification, Web Services Security and Security Assertion Markup Language. He concludes by calling these two methods as 'mature standards' and says that both can be mutually implemented in order to ensure security to XML documents.

Yue-sheng, Meng-tao and Yao (2010), in their paper, starts by talking about the need of specialized security measures for web services, as the existing measures, like SSL, "cannot adapt the request of web service security". They talk about implementing web services security using XML Signatures and XML Encryption. They state that both together can be used for ensuring security of web services, as they integrate into the XML environment of web services. They talk in detail about how signature verification is done and also how encryption is done, after comparing each with the traditional approaches. They conclude by saying that web services are based on SOAP and WSDL standard agreement, which are basic XML formats and also say that "XML signature and encryption technology has met the Web service security needs, which can be as the base of secure Web services."

Bertino, Carminati and Ferrari (2001), in their paper, start by giving an idea of XML security and the three main issues like authenticity, integrity and confidentiality. They mainly talk about the need and development of XML Signature, XML Encryption and access control mechanisms, together with explaining the syntax and structure of XML Signature and XML Encryption. They also discuss about the signature validation mechanism. While discussing about access control, they illustrate using Author-X, "a comprehensive Java-based system for the protection of XML documents". They give an overview of the Author-X architecture, as seen in figure 7, which is used for specification, administration and enforcement of access control policies for XML documents. They also talk about Secure XML Publishing, as seen in figure 8, which involves that Publishers of XML documents should satisfy the authenticity and confidentiality that can be ensured through the use of digital signatures and hashing techniques. They conclude by saying that XML security is still a new developing area where there are many issues that have not yet been addressed like "development of a flexible XML standard for encryption, development of an infrastructure for certifying and evaluating XML subject credentials and the investigation of the relationships between subject credentials and conventional certificate management approaches, such as for instance X.509".

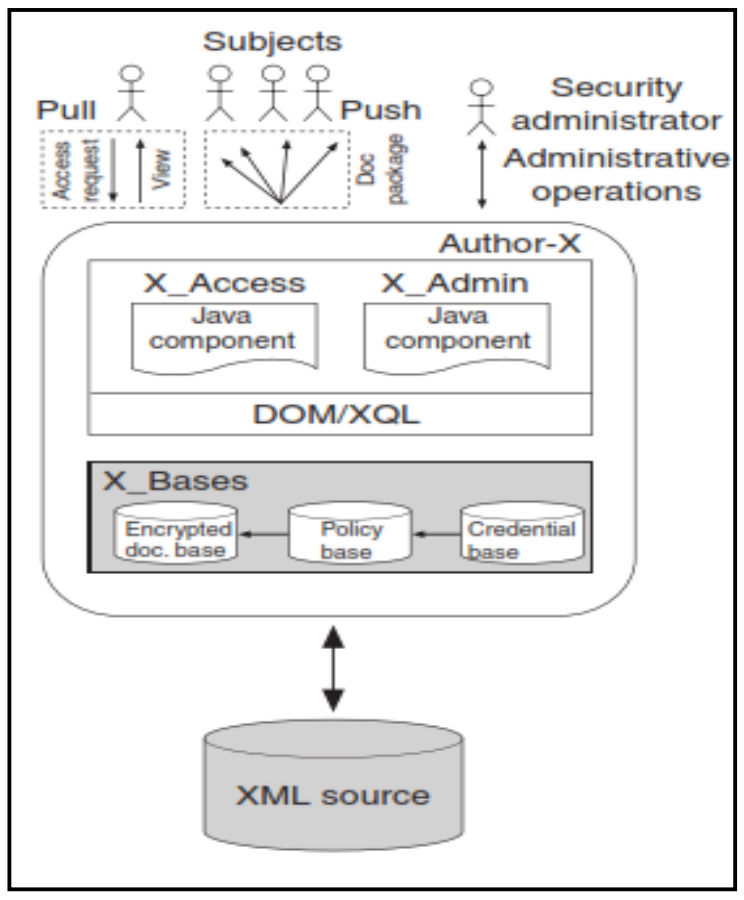

Fig. 7 Architecture of Author-X
(Bertino, Carminati and Ferrari, 2001)

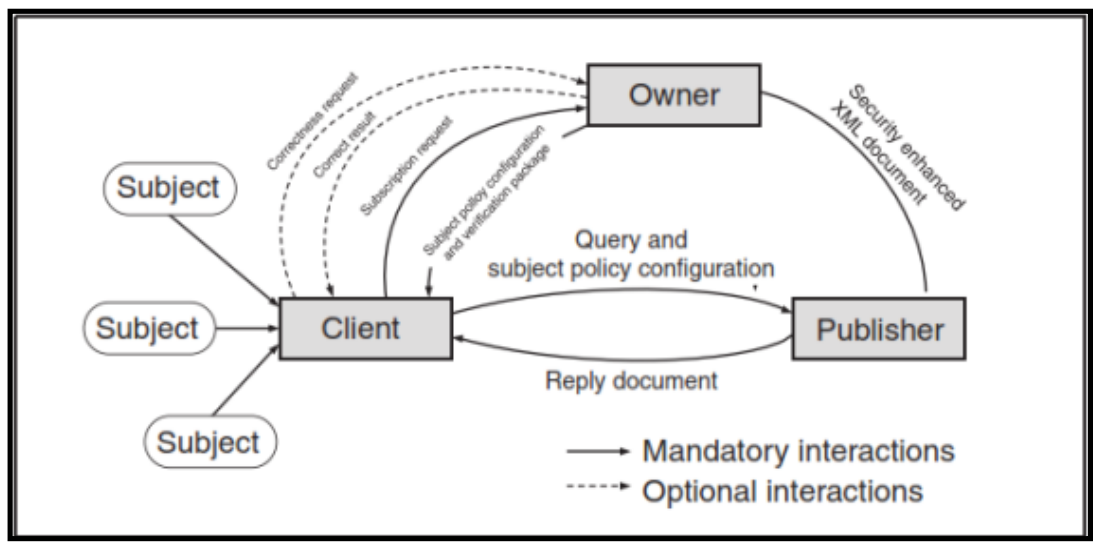

Fig. 8 Secure Publishing Architecture
(Bertino, Carminati and Ferrari, 2001)

Nordbotten (2009), in his paper, presents an overview of the current XML security standards. He says that all these standards together "provide a flexible framework for fulfilling basic security requirements such as confidentiality, integrity, and authentication, as well as more complex requirements such as non-repudiation, authorization, and federated identities". He talks about various XML Security standards and the security measures they provide like XML Signature for integrity, authentication and non-repudiation, XML Encryption for confidentiality and XML Key Management Specification (XKMS) for employing XML Signature and XML Encryption in a scalable manner. With regards to web service security, he discusses about WS-Security for securing integrity and confidentiality of SOAP messages, the Web Services Policy, WS-Trust for augmenting WS-Security and WS-Policy and WS-SecureConversation for secure exchange of multiple messages. He also talks about various security markup languages like eXtensible Access Markup Language (XACML) for defining access control policies using XML and Security Assertion Markup Language (SAML) for defining security assertions (i.e. set of statements made by asserting party that a relying party may understand) in XML.

Yue-sheng, Bao-jian and Wu (2009) talks about the need of XML Signatures inspite of the traditional security technologies like SSL while also stressing on the limitations of SSL like its incapability to encrypt partial document and also its guarantee of only Point-to-Point security. They also explain in detail the XML Signature syntax and processing and compares it with traditional digital signatures. Its application in web services security is explained using a Java implementation of XML signature in the Web services security.

Sun and Li (2005) and Han, Park and Lim (2011) stress the importance of XML Signature and benefits of using them. Sun and Li introduces a concept of undeniable XML signature using undeniable RSA encryption. In normal XML signatures, the verifier need not require the signer's validation for verifying the signed document. This, at times, can be harmful as the signer can deny the signing of document. With the help of undeniable signatures, the verifier requires the signer's verification for verification of the document. Han, Park and Lim discusses the common e-commerce frameworks and the requirement of security in them. They also discuss in detail the XML Signature, its types, its structure, signature verification and application in the e-business environment.

Takase and Uramoto (2002) and Knap and Mlynkova (2009) proposes the idea of using XML digital signatures in the current Internet B2B communications without changing the existing applications, by defining functions for signing and verifying Web Services using SOAP as the message format. Knap and Mlynkova also talk about the various security attacks to XML Signatures that are possible, like XML injection, denial of service and counterfeit XML fragment, and their possible solutions. Takase and Uramoto described a XML digital signature proxy server, implemented as a web service, so that the system is flexible enough to be implemented anywhere.

## 6. Summary

For signing or encryption of an XML document, we need a Digital ID also known as a Digital Certificate. XML Encryption is used to ensure that only the intended receiver is able to understand the message and XML Signature is used to ensure that the message is received by the receiver in the same state as it was sent without any modifications (integrity).

XML Encryption ensures user authentication, authorization, integrity and confidentiality. However, it is practically unimportant to encrypt a data if the same data is to be sent to a group of people. For this reason, using XML Signature still ensures the data integrity along with user authentication.

1. Authentication (Bertino, Carminati and Ferrari, 2001; Han, Park and Lim, 2011; Singhal, Winograd and Scarfone, 2007):
   While implementing XML Encryption, anyone can encrypt the document using public key as it is freely available. But only the signer can authenticate that it is actually done by the right person.
2. Verification (Singhal, Winograd and Scarfone, 2007):
   Encryption and signature both provides verification of the encrypted/signed document as for encryption, you need the private key of the receiver to decrypt and for signature, you need the public key of the sender to verify the rightful sender.
3. Integrity (Bertino, Carminati and Ferrari, 2001; Han, Park and Lim, 2011; Singhal, Winograd and Scarfone, 2007):
   Both encryption and signature guarantees the integrity of the secured document as no one can manipulate the document once it is signed/encrypted.
4. Confidentiality (Bertino, Carminati and Ferrari, 2001; Han, Park and Lim, 2011; Singhal, Winograd and Scarfone, 2007):
   XML Signature does not guarantee confidentiality of the data as the data is still in plain text while in XML Encryption, the data that is displayed is the cipher (or encrypted) text.
5. Non-Repudiation (Han, Park and Lim, 2011; Nordbotten, 2009; Singhal, Winograd and Scarfone, 2007):
   XML Signature provides for non-repudiation of document as the sender or signer cannot deny that he/she has not signed the document, due to the use of keys for signing.
   We can use a combination of both for providing security in web services. Two approaches can be done: first signing and then encryption or first encryption and then signing.
6. First sign and then encryption
   This ensures that the signature remains protected but at the same time involves a greater overhead of decrypting the message all the time to verify the signature. In this method, the encryption algorithm can also be changed without affecting the signature.
7. First encryption and then sign
   In this method, we will immediately come to know if the data has been tampered with, but the sender identity is disclosed as the decryption key is sent to the public.

**Table 1** Security Measures by XML Signature and XML Encryption

| | Authentication | Authorization | Verification | Integrity | Confidentiality | Non-Repudiation |
|---|---|---|---|---|---|---|
| **XML Signature** | Yes | - | Yes | Yes | - | Yes |
| **XML Encryption** | - | Yes | Yes | Yes | Yes | - |
| **Papers** | Bertino, Carminati and Ferrari, 2001; Han, Park and Lim, 2011; Singhal, Winograd and Scarfone, 2007 | Bertino, Carminati and Ferrari, 2001; Singhal, Winograd and Scarfone, 2007 | Singhal, Winograd and Scarfone, 2007 | Bertino, Carminati and Ferrari, 2001; Han, Park and Lim, 2011; Singhal, Winograd and Scarfone, 2007 | Bertino, Carminati and Ferrari, 2001; Han, Park and Lim, 2011; Singhal, Winograd and Scarfone, 2007 | Han, Park and Lim, 2011; Nordbotte, 2009; Singhal, Winograd and Scarfone, 2007 |